# Electrically Driven Single Electron Spin Resonance in a Slanting Zeeman Field

M. Pioro-Ladrière[1], T. Obata[1], Y. Tokura[1,2], Y.-S. Shin[1], T. Kubo[1], K. Yoshida[1], T. Taniyama[3,4] and S. Tarucha[1,5]

[1]*Quantum Spin Information Project, ICORP, Japan Science and Technology Agency, Atsugi-shi,243-0198, Japan.*

[2]*NTT Basic Research Laboratories, NTT Corporation, Atsugi-shi, 243-0198, Japan.*

[3]*Materials and Structures Laboratory, Tokyo Institute of Technology, 4259 Nagatsuta, Yokohama, 226-8503, Japan.*

[4]*PRESTO, Japan Science and Technology Agency, 4-1-8 Honcho Kawaguchi, Saitama 332-0012, Japan.*

[5]*Department of Applied Physics, University of Tokyo, Hongo, Bunkyo-ku, Tokyo, 113-8656, Japan.*

**The rapidly risingfields of spintronicsand quantum information science have led to a strong interest in developing the ability to coherently manipulate electron spins[1].Electron spin resonance[2] (ESR) is apowerfultechniqueto manipulate spinsthat is commonly achievedby applying an oscillating magnetic field. However, the technique has proven very challenging when addressing individual spins[3-5]. In contrast, by mixing the spin and charge degrees of freedom in a controlled way through engineered non-uniform magnetic fields, electron spincan be manipulated electrically without the needof high-frequency magnetic fields[6,7]. Here we realize electrically-driven addressable spin rotations on two individual electrons by integrating a micron-**

**size ferromagnet to a double quantum dot device. We find that the electrical control and spin selectivity is enabled by the micro-magnet's stray magnetic field which can be tailored to multi-dots architecture. Our results demonstrate the feasibility of manipulating electron spinselectrically in a scalable way.**

Magnetic resonance was recently used to coherently manipulate the spin of a single electron[5] in a semiconductor structure, called a quantum dot[8,9], whose tally of electrons can be tuned one by one, down to a single charge[10,11]. However, producing strong and localized oscillating magnetic fields, which is a necessary step to addressing individual spins, is technically demanding.It involves on-chip coils[5,12], relatively bulky to couple with a single spin, dissipating a significant amount of heat close to the electrons whose temperature must not exceed a few decikelvins. In comparison, strong and local electric fields can be generated by simply exciting a tiny gate electrode nearby the target spin with low-level voltages. For scalability purposes, it is therefore highly desirable to manipulate electron spins with electric fields instead of magnetic fields.

To benefit from the advantages of electrical excitation, a mediating mechanism must be in place to couple the electric fieldto the electron spin, which usually responds only to magnetic fields. Spin-orbit coupling[13,14], hyperfine interaction[15]and g-factor modulation[16] work as the mediating mechanism, which attracts interest for their physical origins but necessitate refinement both in terms of manipulation speed and scalability.Instead we controllablymixthe spin and charge degrees of freedom ina magnetic field gradient[6], very much like the Stern &Gerlach effect[17].This allows for greater flexibility since the method is applicable to any semiconductor material. In addition, the magnetic field profile can be engineered to enablethe selective manipulation of several spins using a single electrode.

Thereby, we demonstrate addressable voltage driven single spin ESR in a magnetic field gradient. Two electrons are confined and spatially separated from each other in a gate-defined double quantum dot[18] (Fig. 1a). The ac electric field, $E_{AC}$, is generated by exciting a nearby gate that couples to both electrons. The magnetic field gradient is obtained by using a ferromagnetic strip that we integrate on top of the double-dot structure. The strip is magnetized uniformly along its hard axis by applying an in-plane magnetic field, $B_0$, stronger than the micro-magnet's saturation field (~ 2 T). In this condition, the resulting stray magnetic field has anout-of-plane component that varies linearly with position, pointing in the upward (or downward) direction to the left (or right) of the quantum dot locations (Fig. 1b). In addition, the inhomogeneityof the in-plane component yields two different quantum dot Zeeman fields $B_{0L}$ and $B_{0R}$ (Fig. 1c).We use this feature to probe each spin separately.

To achieve ESR, we displace periodically the two electronsaround their respective equilibrium positionin the slanting field. In each dot, the spin feels an upward magnetic field when the charge is displaced to the left. Conversely, the electronexperiences a magnetic field pointing in the downward direction when displaced to the right. This effective oscillatory magnetic field induces transitions between each electron spin states(pointing in the direction parallel or anti-parallel to the external field $B_0$) only when the driving frequency, $f$ ,matches the Larmor frequency, $f_0$, of the target spin. The latteris proportional to the corresponding quantum dot Zeeman field ($hf_{0L,R} = g\mu_B B_{0L,R}$ where $h$ is the Planck constant, $g$ the Landé factor and $\mu_B$ the Bohr magneton). By adjusting the frequency, phase andduration of the ac electric field burst used to periodically displace the electrons, arbitrary single spin rotationscan then be realized in each dot through the ESR effect, a prerequisite for realizing the CNOT gate usingthe exchange interaction between neighboring spins[19].

To detect the electrically induced spin flips, weoperate the double-dot in the Pauli spin blockade regimewhere no current flows unlessspin-flips occur in either dot[5,20](Fig. 1D). The blockade arises because of Pauli exclusion: the electron in the right dot cannot tunnel to the left dot becauseits spin points in the same direction as the other electron. The spin blockade regime is identified in the double-dot stability diagram by mappingthe dot current, $I_{dot}$, under large source-drain bias as a function of the left and rightquantum dot gate voltages (Fig. 1e).

We now showthat electric excitation can induce single-electron spin flips. We apply a continuous microwave voltage and followthe stability diagram around the resonance condition (Fig. 2a). The resulting electric field modifies the diagram through a process known as photon-assisted tunneling[21] (PAT).In general, PAT can assist electrons in breaking the spin blockade by, for instance, hopping to an un-blockedspin state, usually energetically inaccessible. For our power level, a clear spin-blockade region remains with leakage current below the noise floor of the experiment (20 fA).

The situation is different at the resonance condition for ESRwhere a finite leakage current now flows in the spin-blockaded region. By resonantly flipping the spin of the electron residing on either dot, the electron on the right dot is now allowed to tunnel to the left dot thereby lifting the spin blockade.The resonantresponse is observed clearly as $B_0$ and $f$are varied for constant $E_{AC}$ (Fig. 2b). Two equally spaced peaks (with spacing $\Delta_Z$ = 13 ± 2mT) in $I_{dot}$ are seenat a frequency proportional to $B_0$. Judging from the amplitudes, we attribute the first (second) peak to spin flips of the electron residing on the left (right) dot. This selective addressing is enabled by the inhomogeneous in-plane stray field profile mentioned above.By slightly modifying the micro-magnet geometry, the frequency selectivity can, in principle, be used to address individual spins in a scalable way (Supplementary note).

The linear dependence of each resonance on the external magnetic field is a key signature of ESR since the Larmor frequency is proportional to $B_0$.[22]From the averaged position of one of the ESR peak obtained over a wider range of magnetic field (Fig. 2c), we determine $g$ = 0.41 ± 0.01, in good agreement for our type of device.Following the peak positionbelow the micro-magnet's saturation field, we have confirmed that $B_{0L,R}$ are smaller than the external field, afeature expected for the stray magnetic field (Supplementary Fig. 1a).

Evidence for spin-charge coupling induced by the slanting field is revealed in the ESR peak height. This gives information on the effective ac magnetic field strength, $B_{AC}$, which is proportional not only to $E_{AC}$ but also to the magnetic field gradient, $b_{SL}$. To estimate $B_{AC}$, we use the non-monotonous response of the peak height with microwave power. As the power level is raised, the peak amplitude initially increases and then saturates past a certain level corresponding to anelectric field $E^*_{AC}$ (inset of Fig. 3). This response results from the interplay between the ESR and fluctuating Overhauser fields. The Overhauser field arises from hyperfine interaction between the electron and nuclei spins of the host material[23]. This interaction shifts theLarmor frequency randomly by an amount $\Delta f_0 = g\mu_B B_N / h$where $B_N$ is the amplitude of the nuclear field fluctuations.For $B_{AC} > B_N$, power broadening washes out the fluctuations. Every time an electron blocks the transport by spin-blockade, the ESR field flips its spin and the current flow is therefore saturated. For $B_{AC} < B_N$the resonance condition is only met occasionally. Fewer electron spins are flipped per unit of time and the current is consequently lower. The saturation occurs at $B_{AC} \sim B_N / 2$.[15,24]

The Overhauser field fluctuations are also responsible for the jitter in the peak position visible in Fig. 2bwhich allows us to extract $B_N$ =2.4mT.Using this result, we estimate $B_{AC}$ to 1 mT at the onset of saturation. Quite remarkably, such magnitude is obtained for microwavepower five hundred times smaller than for magnetically driven ESRwith on-chip coil[5,12]. By operating deeper in the Coulomb blockade region of the stability diagram, fields as strong as 10mT are possible since stronger PAT is required to lift the spin-blockade, yielding a spin-flip time as fast as 20 nsec.The efficiency can further be improved by increasing the micro-magnet thickness and using stronger ferromagnetic materials[7].

In Figure 3, we plot the estimated spin-flip rate (Rabi frequency, $\nu_{Rabi} = g\mu_B B_{AC} / 2h$ )normalized over electric fieldin a magnetic field range above the micro-magnet saturation field. The normalized Rabi frequency does not vary significantly with $B_0$ , as expected since $b_{SL}$ should be constant in this regime.A linear fit through the data suggests that a second fieldalso contributes, on a smaller level, to theeffective ESR field. We attribute the second contribution tothe spin-orbit interaction. The latter gives rise to an intrinsicslanting fieldof slope $2B_0 / l_{so}$ , where $l_{so}$ is the characteristic spin-orbit length[25]. Assuming the micro-magnet stray fieldhas a gradient of opposite sign (pointing downward when the intrinsic spin-orbit field points upward), the spin-orbit interaction works in reducing the total slanting field, a trend observed in the data[26]. To be more quantitative, we derivedthe following expression for the total effective ESRfield strength:

$$B_{AC} = \frac{eE_{AC} l_{orb}}{\Delta}\left( |b_{SL}| - \left|\frac{2B_0}{l_{so}}\right| \right) l_{orb} \quad (1)$$

where $\Delta$ and $l_{orb}$ are the quantum dot's confinement energy andorbital spread. The fit to equation (1) yields $b_{SL}$ ~ 0.8 T/μm, in good agreement with the expected stray magnetic field profile, and $l_{so}$ ~ 58 μm whose magnitude is consistent with recently observed spin-orbit mediated ESR in a similar system[13].The fluctuating nuclei field was shown to also enable electrically driven spin flips[15]and should, like the spin-orbit case, contributes to our ESR signal. However, the weak ESR response observed at low external magnetic fields (where $b_{SL} \sim 0$) implies that the hyperfine effect does not contribute significantly to the effective ESR field (Supplementary Fig. 2).

The spin rotations demonstrated here, in combination with experimentally realized spin read-out[27,28]and tuneable exchange coupling[29,30]fulfil many of the requirements for quantum computing with electron spins in quantum dots[19] using only electric fields. In contrast to previously reported voltage driven ESR mediated by spin-orbit coupling or interaction with nuclei spins, our scheme, which is applicable to any material, does not rely on intrinsic properties which are also responsible for degrading the spin coherence in solid-state systems. The coherence time for our hybridized spin is expected to beas long as 1 msusingcleaner materials[6] such as carbonnanotubes, Si nanowires and SiGeheterostructure. Moreover, micro-magnetsmay simplifythe daunting task of integrating many quantum dots into a multi-qubit quantum register. The independent addressing of the spin in each of the double-dot observed here and in Ref. 15inspiresscalability with the help of micro-magnets. By engineering the stray field profile, a common ESR gate could be used to operate on any spin in the register, simply by matching the driving frequency to a position dependent Zeeman field.

M. P.-L. acknowledges F. H. L. Koppens for advices and I. Mahboob for comments.

Correspondence and requests for materials should be addressed to M. P.-L. (michel@tarucha.jst.go.jp).

**Figure 1:Device and read-out scheme.a**Scanning electron micrograph of a device similar to the one used in the experiment. The Ti/Au gates (light grey) are deposited on top of a GaAs/AlGaAsheterostructure containing a two-dimensional electron gas 90 nm below the surface. The 70 nm thick Cobalt micro-magnet (artificially colored in yellow) has been displaced from its actual location (marked by the dashed lines) for clarity. The micro-magnet is isolated from the gate structure by an 80 nm thick calixarene insulating layer. White arrows indicate the current flow through the two quantum dots (dotted circles) connected in series to source and drain. **b**The magnetization $\vec{M}$ (parallel to in-plane magnetic field $\vec{B}_0$) produces across the quantum dots (only one is shown in blue for clarity) a transverse stray magnetic field (red arrows) of the slanting form: $\vec{B}_x = b_{SL} z\hat{x}$. The periodicaldisplacement of the electron's wavefunctionis driven by the ac gate voltage $V_{AC}$ . **c** Simulated profiles of the in-plane field shift, $\delta B_0 = B_z - B_0$ , and gradient of the transverse stray field, $b_{SL} = \partial_z B_x$ , near the double-dot location. The origin is taken at the center of the micro-magnet. For slight misalignment, the in-plane field produces two different quantum dot Zeeman fields $B_{0L}$ and $B_{0R}$ . **d**The sequential flow of electrons at spinblockade. Starting with the $(N_L, N_R) = (1,0)$ charge state, an electron tunnels from the sourceto form the $(1,1)$state. The electron in the right dot cannot tunnel to the left dot because of Pauli exclusion principle and transport is blocked.With ESR, electron on the right dot cantunnel to the left dot to form the $(2,0)$ state. One of the two electrons thentunnels out to the drain to complete the cycle $(1,0) \rightarrow (1,1) \rightarrow (2,0) \rightarrow (1,0)$ , yielding a finite leakage current. **e**Stability diagram measured with $V_{AC} = 0$ for source-drain bias $V_{SD}$ = 1.4 mV at $B_0$ = 2 T. In the regions marked by the corresponding charge states $(N_L, N_R)$, transport is quenched due to Coulomb blockade where the left (right) dot holds a fixed number of electrons $N_L$ ($N_R$) tuned by gate voltage $V_L$ ($V_R$). Coulomb blockade is lifted in the triangular regions where current flows except for the spin blockade area (enclosed by orange dotted lines).

**Figure 2: Electrically driven single spin resonance.a**Stability diagrams under continuous electric excitation taken at frequency *f* = 26.5 GHz and power level of -38 dBm. In the left (right) color plots, the Larmor frequency in the left dot is detuned (tuned) to *f* using the external magnetic field ($B_0$ = 4.7 and 4.712 T respectively). **b**(left) Dot current as function of $B_0$ measured after adjusting $V_L$ and $V_R$ to a resonant point of the spin blockade region. (right) Similar scans obtained over a small range of frequencies. The dashed lines are guides to the eye to indicatethe linear dependence of the two ESR peaks associated to the left and right dot. The power level is adjusted at each frequency to keep $E_{AC}$ constant. **c**The two ESR peaks obtained over a wide range of magnetic field. Each trace is offset by an amount proportional to the corresponding frequency. Inset: Position of the left dot peak. Each data point is obtained by averaging the peak position over 5 magnetic field sweeps. Error bars are smaller than the symbols. Dotted red line is a linear fit yielding $g = 0.41 \pm 0.01$.

**Figure 3:Dependence of the Rabi frequency on external field.** Each data point is obtained after performing a power dependence of the ESR peak height (associated to the left dot) to extract the value of the electric field at saturation, $E^*_{AC}$, and corresponding $B_{AC}$.The inset shows such power dependence obtainedat $B_0$ = 2.14 T. The dot current is normalized over the saturation current $I_0$, which also depends on power due to PAT process (Supplementary Fig. 2). The red line is a fit

to equation (1) assuming harmonic confinement with estimated energy $\eta\omega_0$ ~ 1 meV ( $\Delta = \eta\omega_0$ , $l_{dot} = \sqrt{\eta / m\omega_0}$ where $m$ is the effective electron's mass, equal to 0.067 times the free electron mass) and $g$ = 0.41.

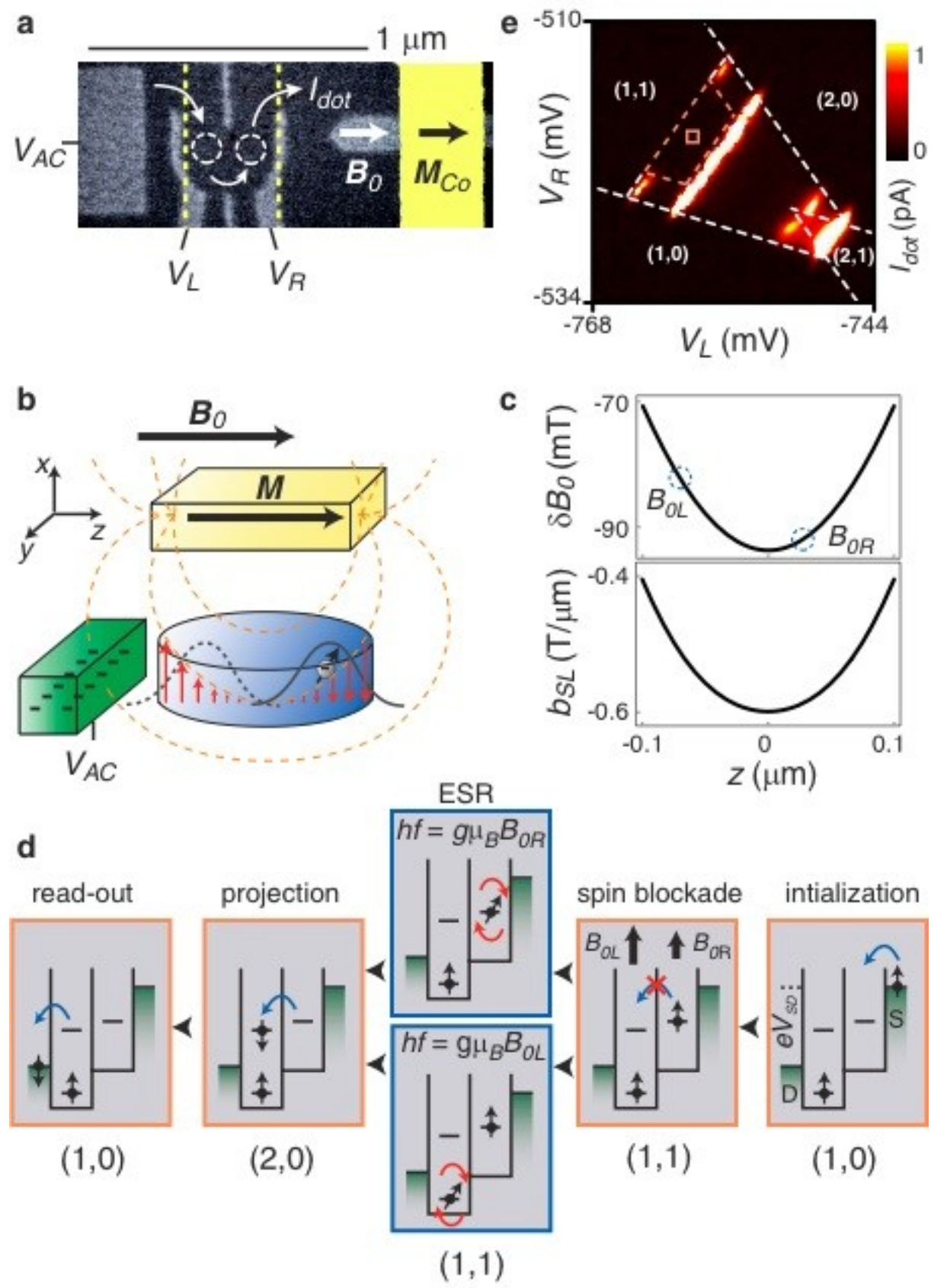

Pioro-Ladrière et al. Figure 1

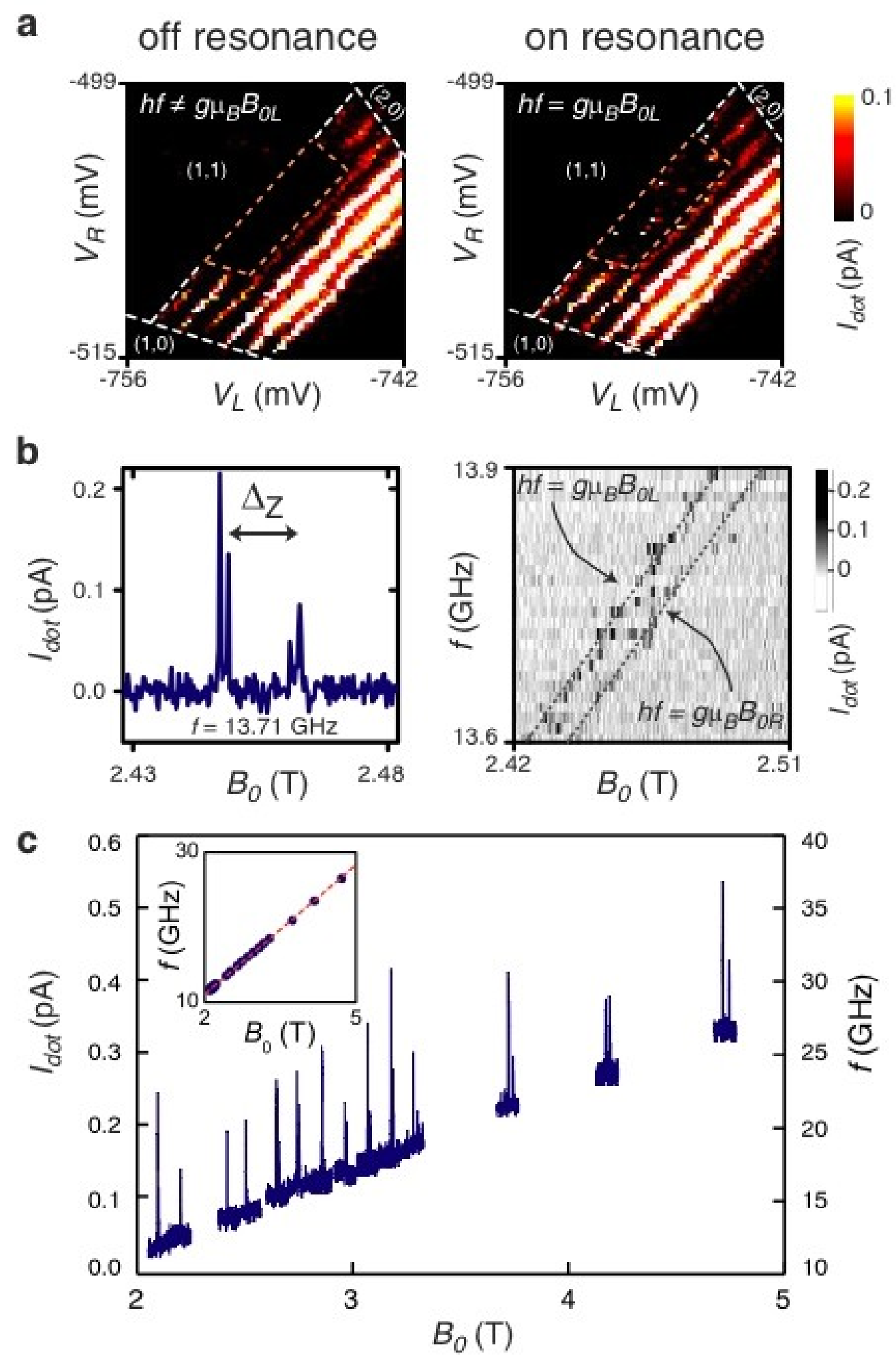

Pioro-Ladrière et al. Figure 2

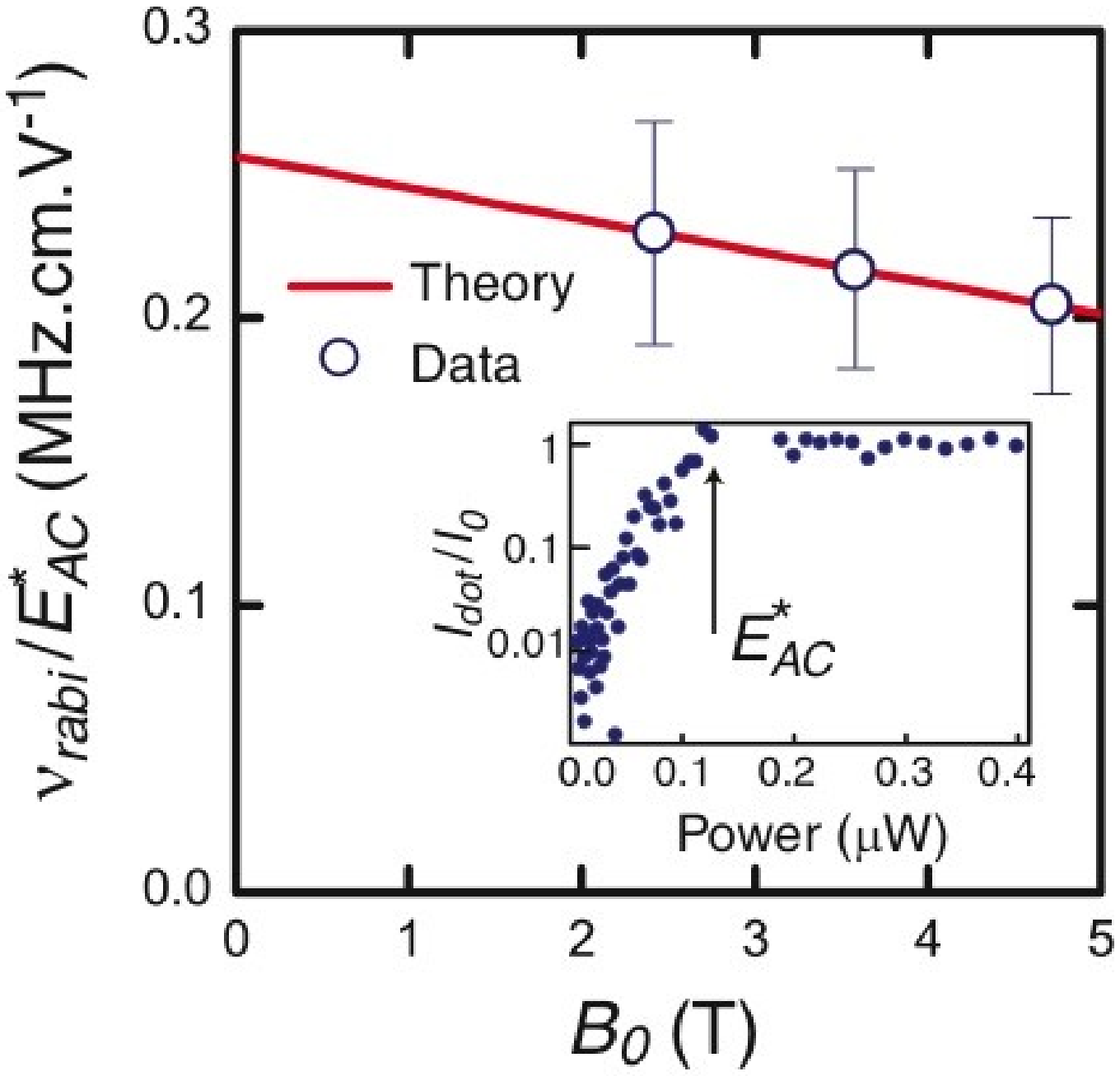

Pioro-Ladrière et al. Figure 3

# Supplementary Material:

# Electrically Driven Single Electron Spin Resonance in a Slanting Zeeman Field

M.Pioro-Ladrière, T. Obata, Y.Tokura, Y.-S. Shin, T. Kubo, K. Yoshida, T.Taniyama, S. Tarucha

## A. Supplementary Materials and Methods

The GaAs/AlGaAsheterostructure from which the sample was made was purchased from Sumitomo Electric. The two-dimensional electron gas (2DEG) has a mobility of 1 x $10^2$ $m^2V^{-1}s^{-1}$ and an electron density of 3 x $10^{15}$ $m^{-2}$ measured at 1.4 K. The device used in this experiment suffered initially from a large level of telegraphic noise, associated to the switching of a few background charges. This made the quantum dot behaviour extremely unstable. To improve the charge stability, the device was first cooled down under a positive bias of +0.5 V applied on the quantum dot gates. A negative bias of -2 V was then applied on the micro-magnet once the device had reachedthe base temperature.

The occupation of the double-dot by only two electrons was confirmed by first opening the inter-dot barrier (using P gate) to form a single dot and then by opening the barriers separating the dot to the source and drain reservoirs (using T gate). Under these conditions, no extra Coulomb blockade peaks appeared in the region $N_L + N_R = 0$ of the double-dot stability diagram.

The stray magnetic field produced by the ferromagnetic strip is calculated numerically using the Mathematica© package Radia, available at http://www.esrf.fr, assuming a uniform magnetization and using the saturation magnetization of Cobalt ( $\mu_0 M_{Co}$ = 1.8 T). The strip is 5 μm long.

The microwave signal is applied to the ESR gate using a commercial microwave source (Agilent 8360B). The ESR gate is part of an on-chip co-planar waveguide (CPW) which is wire bonded to the sample holder's alumina CPW. The latter is connected to the high-frequency line of the dilution refrigerator using a microwave bead. These precautions are taken to minimize loss by improving impedance mismatches.

The measurements were performed in an Oxford Instrument Kelvinox 100 dilution refrigerator operating at a base temperature of 40 mK.

## B. Supplementary Note

**B.1. Engineering the stray magnetic field to address several spins in a quantum dot array.**

We show in this section how several spins could be addressed using a common ESR gate electrode by slightly modifying the micro-magnet geometry. As in (*14*), we assume a linear array of gate-defined quantum dots, each holding a single electron (Fig. S3A). To perform single spin operation on any spin of the array, we employ the concept demonstrated in the main text, i.e. frequency selective voltage driven ESR. To achieve this, two large ferromagnetic strips are deposited on top of the quantum dot array. The separation between the two strips has two purposes. The first one is to produce a transverse slanting magnetic field at each dot location (Fig. S3B). The second one is to provide a strong electric ac electric field, which is approximately equal to the applied ac voltage between the two strips over the separation. By adjusting the frequency, phase and duration of the electric field, any single spin rotation can then be realized.

The frequency selectivity is achieved by tapering the strip separation, yielding different a Zeeman field $B_{0i} = B_0 + \delta B_{0i}$ for each dot location *i*. This is possible because the stray field induced shift, $\delta B_{0i}$, is a function of the separation $d_i$ in the taper. For 100% addressability efficiency, the difference in Larmor frequency, $\Delta f$ , between two adjacent spins must be greater than the inverse of the intrinsic spin coherence time $T_2$ . In Fig. S3C, we present simulation results for an array of 60 spins using realistic device parameters. The chosen taper yields $\Delta f \approx 10$ MHz, which is much bigger than $T_2^{-1}$ for nuclei spin free systems (*10*). The field gradient, $b_{SL}$ , necessary for fast single spin rotation is above 1 T/μm throughout the array.

The hybridization between the spin and charge degrees of freedom induced by the slanting magnetic field slightly modifies the exchange interaction between neighboring spins. However, as demonstrated in (*10*), the CNOT gate can still be realized in such hybrid systems. Combined with the single spin addressability explained above, the proposed architecture brings all-electrical universal control of electron spins within reach.

**C. Supplementary Figures**

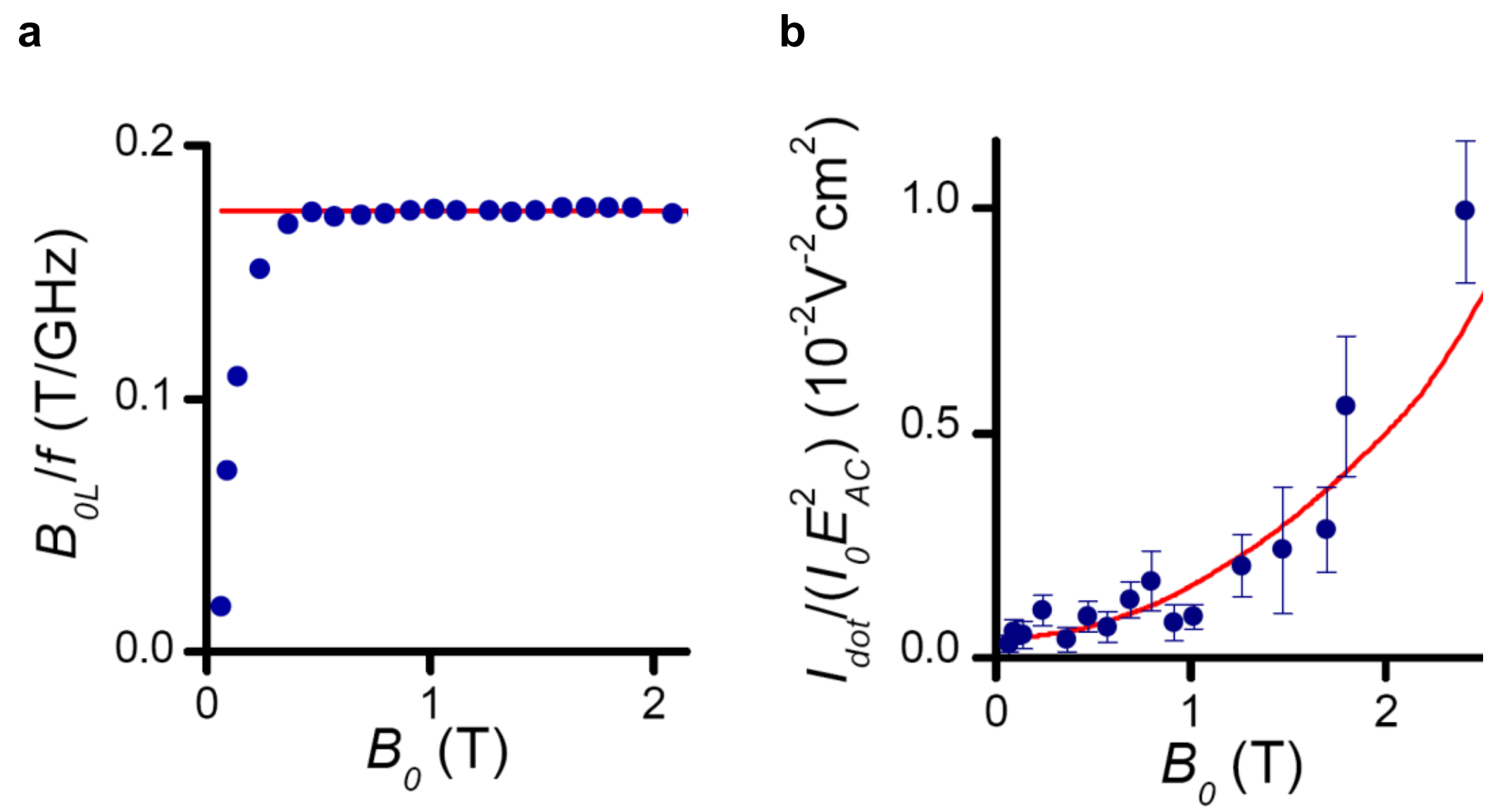

**Supplementary Figure 1.a**Evolution of the ESR peak position (associated to the left dot) as the micro-magnet is magnetized by the external field $B_0$. Plotted is the value of total quantum dot Zeeman field $B_{0L} = B_0 + \delta B_{0L}$ normalized over the resonant frequency *f*. A constant in-plane shift $\delta B_{0L}$ = -60 mT is assumed. The red line is obtained using $g = 0.41$. The deviation indicates variation in the stray Zeeman field by the micro-magnet magnetization process. **b** Corresponding peak height, normalized over the saturation current $I_0$ and the square of the ac electric field $E_{AC}$. The error bar is dominated by the uncertainty in $E_{AC}$ and $I_0$. The red curve is a parabolic fit with small positive offset. The ESR peaks are obtained at microwave power below the onset of ESR saturation where the height is approximately proportional to $b_{SL}^2 E_{AC}^2$. The parabolic increase is consistent with a linear magnetization curve since $b_{SL} \propto M$. The residual peak at zero external field can be attributed to ESR mediated by the hyperfine field as demonstrated in (*9*).

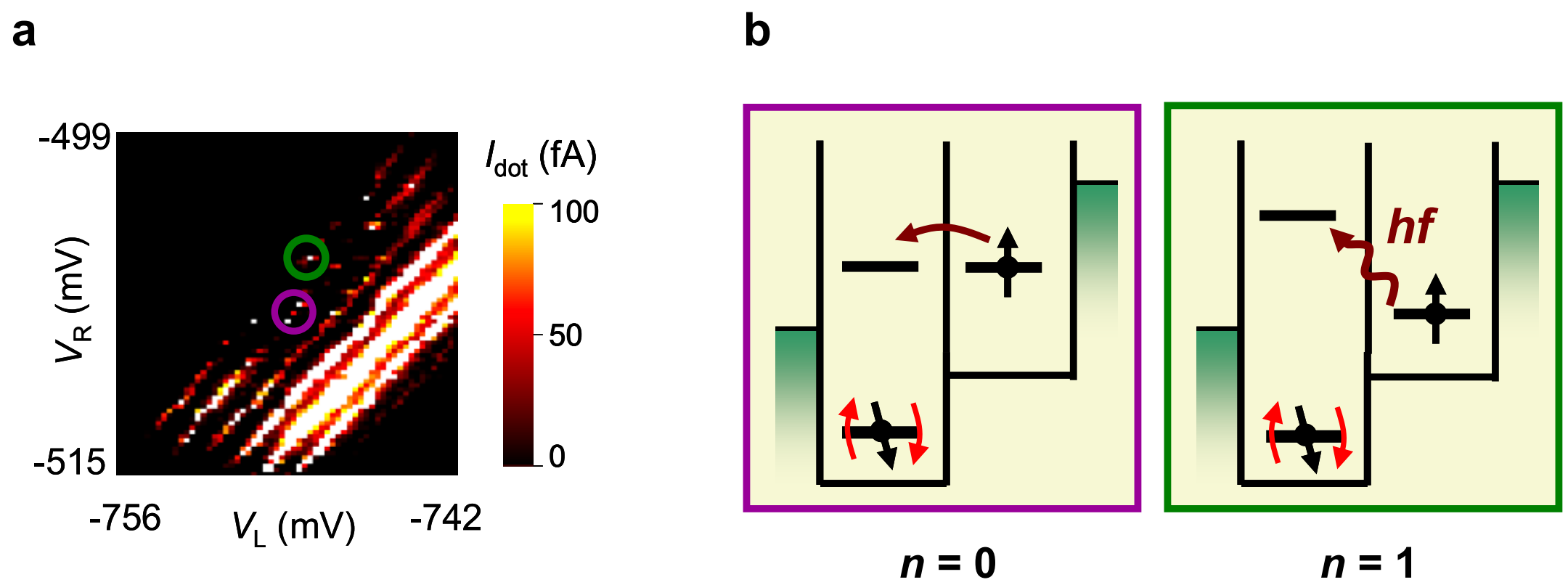
a
b
-499
-515
-756
-742
$V_R$ (mV)
$V_L$ (mV)
$I_{dot}$ (fA)
100
50
0
hf
*n* = 0
*n* = 1

**c**

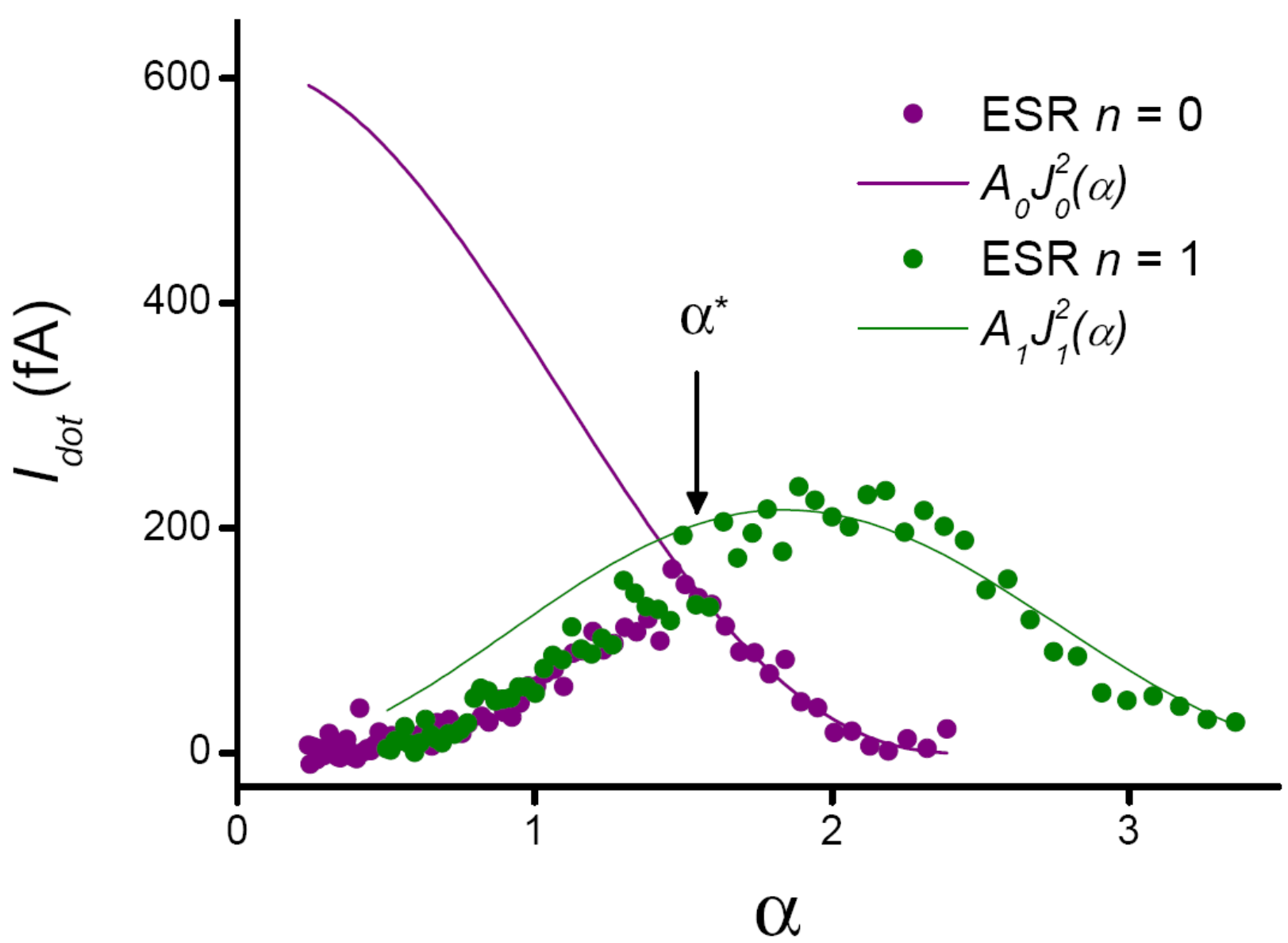

**Supplementary Figure 2.a** Stability diagram obtained at the resonance condition for the left dot. Parameters are the same as in Fig. 2A. The circles indicate typical operation points used for power dependence. **b**Energy diagrams corresponding to the two operation pointin**a**. In both diagrams, ESR lifts off the spin blockade. For the $n=0$ ($n=1$) case, the electron in the right dot tunnels elastically (inelastically) to the left dot. The inelastic transition is enabled by the ac potential drop across the barrier separating the two dots, $V_b$, through photon assisted tunnelling (PAT). The label $n$ indicates the number of microwave photon absorbed in the tunnelling process. For each case, the tunnelling rate is proportional to the square of the corresponding Bessel function, $J_n(\alpha)$, where $\alpha = eV_b/(hf)$ characterize the strength of PAT (*16*). **c** ESR peak height as function of parameter $\alpha$ (obtained at frequency $f$ = 20.1 GHz). The purple (green) data points are taken at the $n=0$ ($n=1$) operation point by varying the microwave source power. Each data point is obtained by extracting the maximum dot current, $I_{dot}$, from a single external magnetic field sweep across the resonance condition. The solid lines are fits of the saturated region ($\alpha > \alpha^*$) to the expected $J_n^2(\alpha)$ behaviour for the cases $n = 0,1$. The two independent fits yield similar amplitudes $A_0$= 611 fA and $A_1$ = 634 fA. The saturation curve shown in Fig. 3 is obtained by dividing the $n=0$ data by the power-dependent saturation current $I_0 \equiv A_0 J_0^2(\alpha)$. The saturation field, $E_{AC}^*$, is estimated using the value of $\alpha^*$ and the simple relation $V_b = E_{AC}/d$ where $d$ is the distance between the two dots, taken as 100 nm.

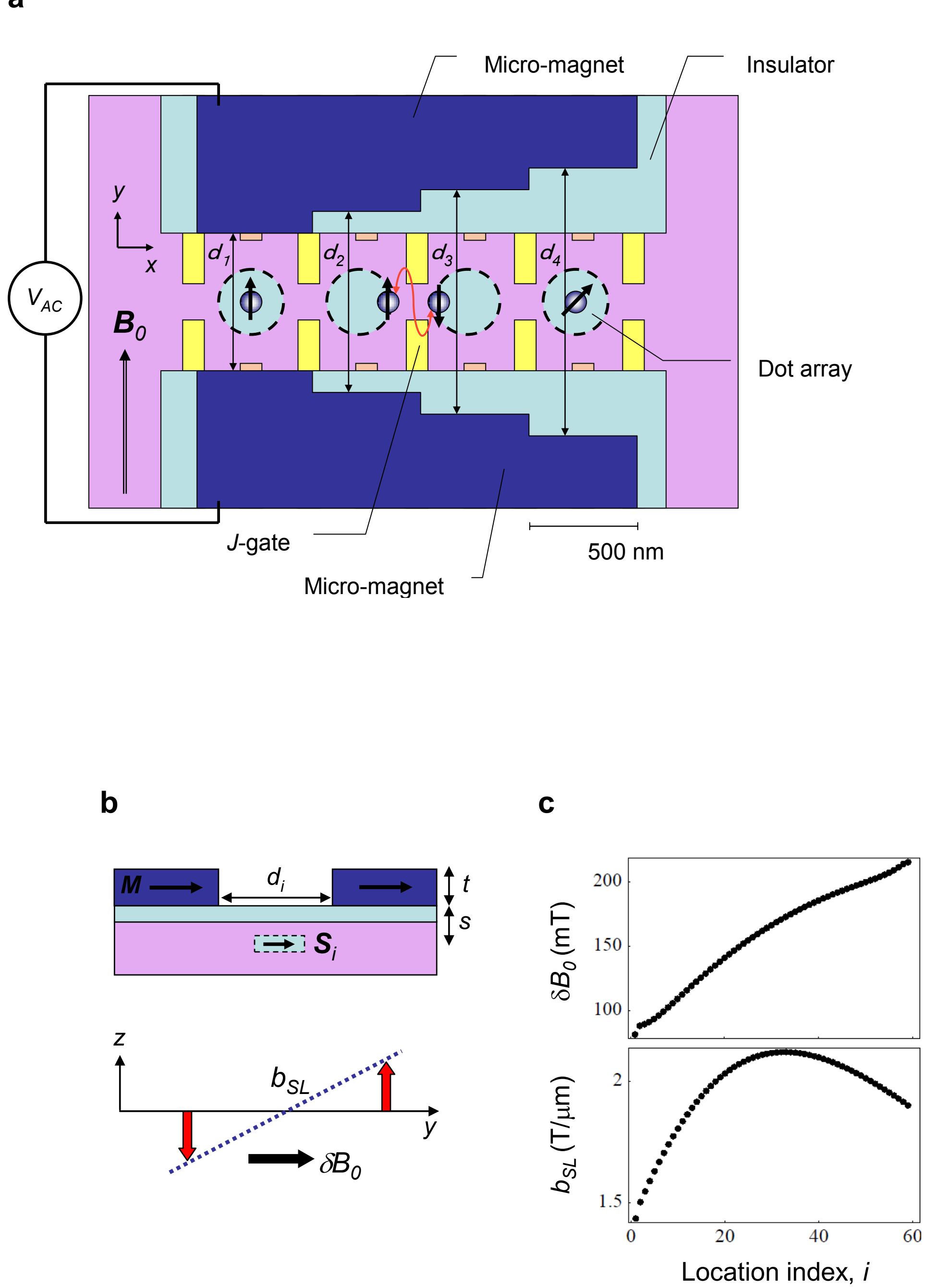

**Supplementary Figure 3.a** Quantum dot array with specially designed micro-magnet assembly. The assembly consists of two ferromagnet strips with a tapered separation $d_i$. **b** Cross-sectional view at the *i*-th quantum dot location. The two strips are magnetized uniformly along the in-plane direction $y$ using the external Zeeman field $B_0$. The stray field transverse component (*z*-component) at the dot location (shown by red arrows) is of the slanting form with gradient $b_{SL}$. The in-plane component (shown in black, *y*-component) shifts the Zeeman field by an amount $\delta B_0$. **c** Numerical simulation of the in-plane shift and transverse gradient calculated at each spin location. The array consists of 60 quantum dots.

The separation increases in increment, $\Delta d = d_{i+1} - d_i$, of 3 nm with $d_{i=1} = 70$ nm. The micro-magnet thickness, $t$, is 150 nm. A distance $s$ of 100 nm from the quantum dot plane to the insulating layer top surface is assumed. The J-gates are used for two-spin SWAP operation as explained in (*14*) and demonstrated in (*25*).